\documentclass[aps, prb, amssymb, amsmath, superscriptaddress, 
twocolumn] {revtex4-2}

\usepackage{chngcntr}

\usepackage{graphicx}
\usepackage{color}
\usepackage{amsmath}
\usepackage{enumitem}
\usepackage{amssymb}
\usepackage[hidelinks]{hyperref}
\usepackage{cancel}
\usepackage{ulem}
\usepackage{multirow}

\newcommand{\be}{\begin{equation}}
	\newcommand{\ee}{\end{equation}}

\newcommand{\bea}{\begin{eqnarray}}
	\newcommand{\eea}{\end{eqnarray}}

\renewcommand{\vec}[1]{{\boldsymbol #1}}

\makeatother

\begin{document}
	\title{ A model of non-Fermi liquid with power law resistivity: strange metal with a not-so-strange origin.}
	
\author{ Patrick A. Lee}
\affiliation{
Department of Physics, Massachusetts Institute of Technology, Cambridge, MA, USA
}

\begin{abstract}
We construct a model which exhibits resistivity going as a power law in temperature $T$, as $T^\alpha$ down to the lowest temperature. There is no residual resistivity because we assume the absence of disorder and momentum relaxation is due to umklapp scattering. Our model consists of a quantum spin liquid state with spinon Fermi surface and a hole Fermi surface made out of doped holes. The key ingredient is a set of singular $2k_F$  modes living on a ring in momentum space. Depending on parameters, $\alpha$ may be unity (strange metal) or even smaller. The model may be applicable to a doped organic compound, which has been found to exhibit  linear T resistivity. We conclude that it is possible to obtain strange metal behavior starting with a model which is not so strange.

\end{abstract} 

\maketitle

\section{Introduction}

The notion of strange metals emerged out of transport measurements from the early days of cuprate superconductivity and  has been applied broadly over the years to instances where linear in temperature resistivity is observed, whether it is low temperature, high temperature  and with or without a finite residual resistivity due to disorder. In this broad sense, strange metal behavior has been seen in many materials and has attracted a great deal of attention from the community, as summarized nicely in  recent reviews ~\cite{phillips2022stranger,hartnoll2022colloquium}. The strange metal is usually associated with the violation of Landau's theory of Fermi liquid, and believed to be driven by strong correlation physics. In this paper we will restrict the use of the term strange metal to low temperatures and in the absence of  disorder. The high temperature linear resistivity anomaly is associated with violations of the Mott-Ioffe-Regel limit and clearly requires a separate set of physical input.~\cite{hartnoll2022colloquium} 
On the opposite end, while there are examples  where linear T resistivity survives to low temperatures, 
in many cases the linear regime does not extend above the residual resistivity beyond  a value  comparable to the residual resistivity itself. Examples of this behavior include the overdoped~\cite{putzke2021reduced} and electron doped cuprates~\cite{greene2020strange}. While this phenomenon is not understood and is of great interest, the effect of disorder is likely to be strongly relevant and  a different set of explanations may be required. Furthermore, as a matter of principle, Landau's Fermi liquid theory refers to the clean case. In this paper we will not discuss models with  disorder~\cite{rosch2000magnetotransport,  maslov2011resistivity,  
  cha2020linear,patel2023universal}. We will focus on the situation where a power law resistivity extrapolates to a small residual value, so that the power extends over a  range much larger than the residual resistivity and disorder may be considered unimportant. Our goal is to produce a model which produces a power law resistivity $T^\alpha$ with $\alpha<2$ which is valid down to zero temperature and which include the linear resistivity as a special case $\alpha=1$. As explained below (see also ~\cite{maslov2011resistivity}) there are barriers towards accomplishing this goal which may explain the paucity of such models. While our model is unlikely to be applicable to cuprate superconductors, it may find application in a doped organic compound.~\cite{suzuki2022mott}. 

Why is it difficult to construct such a model? In the absence of disorder, the total momentum is conserved under scattering unless umklapp scattering is allowed. Therefore in the absence of both disorder and umklapp scattering the conductivity $\sigma(\omega)$ is proportional to $\delta(\omega)$ which is incompatible with any power law function $\omega^{-\alpha}$. Hence the coefficient of $\delta(\omega)$ must be zero. This case  was considered by Else and Senthil~\cite{else2021strange} who show that a power law resistivity requires the divergence of a certain kind of fluctuations in order to kill the delta function. We shall not pursue this route here and instead consider  models where umklapp scattering is allowed down to $T=0$. Such models immediately rule out scattering from critical modes at $q=0$ such as those from ferromagnetic or nematic order or from emergent gauge fields.~\cite{maslov2011resistivity, lee2021low} This leaves soft modes  which are associated with critical point involving ordering at a finite momentum such as anti-ferromagnets or charge density waves. Such scatterings are limited to "hot spots" on the Fermi surface. Without scatterings which rapidly equilibrate the other momentum states on the Fermi surface, the resistivity is dominated by scattering rates away from the hot spots.~\cite{hlubina1995resistivity, rosch2000magnetotransport} This problem can be  brought under control with the introduction of disorder scattering.~\cite{ rosch2000magnetotransport, maslov2011resistivity, lee2021low} This indeed gives rise to a linear $T$ regime with a coefficient which is independent of the disorder. However, the price one pays is that this regime is limited to an increase of resistivity which is equal or less than the residual resistivity. Hence the zero disorder limit cannot be taken without losing the linear $T$ regime.~\cite{rosch2000magnetotransport,lee2021low}

It turns out that these difficulties can be circumvented in a model of doping holes into a quantum spin liquid with a spinon Fermi surface. The spinon Fermi surface has a singular self energy and violates the Landau criterion for quasi-particles. On a triangular lattice umklapp scattering is possible if the density of doped holes is large enough. We shall show that in this case the hot spot becomes a hot region on the Fermi surface, which may even cover the entire Fermi surface. So the bottleneck problem mentioned earlier does not arise and the power law resistivity survives to low temperature. We also emphasize that if the power law $\alpha$ is equal to or less than unity, the Landau criterion is violated and quasi-particles as defined by Landau do not exist. Nevertheless, it was shown long ago by Prange and Kadanoff~\cite{prange1964transport} that even without Landau's quasi-particles, a Boltzmann equation that describes transport properties can be derived. Taking advantage of their insight, the resistivity can be calculated in a simple way even in the "non-Fermi liquid" case.  A quantum spin liquid with  a spinon Fermi surface is  an exotic state of matter, but it is no longer considered very strange. In fact, these state may be realized in certain organic compounds~\cite{zhou2017quantum} and in monolayer 1T-TaSe$_2$~\cite{ruan2021evidence}. In any case, it is useful to have an example which can exhibit strange metal behavior based on a model that is not so strange.

\begin{figure}[h]
\includegraphics[width=0.45\textwidth]{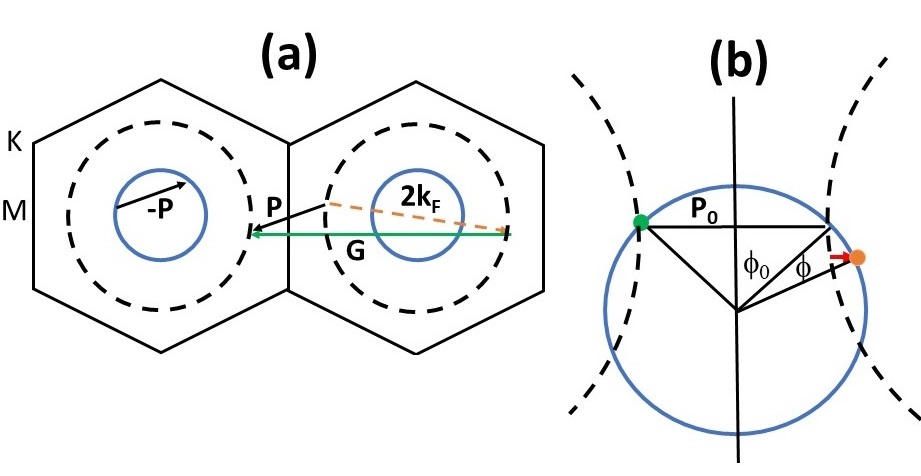}
\caption{(a) Dashed circle shows the spinon Fermi surface in the hexagonal Brillouin zone for a triangular lattice. Solid blue circle shows the hole Fermi surface holding $p$ doped holes. The two Fermi surfaces can exchange momentum $\vec{P}$ via an umklapp process where $\vec{P}$ is the sum of a $2k_F$ vector (dashed  orange line) and a reciprocal lattice vector $ \vec{G}$ (green line) (b) An expanded view of the region between the two spinon Fermi surfaces shown in (a). $\vec{P_0}$ is the shortest umklapp vector connecting the two spinon Fermi surfaces. The hole Fermi surface is positioned so that its Fermi surface is spanned by $\vec{P_0}$. An initial hole state (green dot) at angle $-\phi_0$ is scattered to a final hole state (orange dot) at angle $\phi_0+\phi$ in an umklapp process. The short red arrow denote the vector $\Delta \vec{q}$ (defined above Eq. \ref{Pi}) which measures the deviation of the collective mode momentum $\vec{Q}$ from $2k_F$  } \label{fig1}
\end{figure}

\section{The model}
\label{sec:The model}
We consider  a Hubbard model on the triangular lattice with  nearest neighbor hopping $t$ and onsite $U$. We assume $U/t$ is such that we are on the insulator side of the Mott transition, but not deep inside, so the charge gap is finite but relatively small.  $S=1/2$ local moments are formed on the sites and we assume that they do not order but form a spin liquid state. We further assume that the electrons have fractionalized into fermionic spinons carrying $S=1/2$ but no charge, and relativistic bosonic chargons. Both are coupled to emergent U(1) gauge fields.~\cite{lee2005u} This state was proposed to characterized the organic spin liquids compounds, the ET and dmit salts.~\cite{zhou2017quantum} Recently the ET salt is found to have a spin gap below a phase transition at 6K, so if there is a Fermi surface, at leasst part of it is gapped out at low temperatures.~\cite{miksch2021gapped} On the other hand the dmit salt does not show this transition or gap and  continues to be a good candidate. For ET salts, we assume the spinon Fermi surface is a close-by competing state. More recently the spinon Fermi surface state was also proposed to be realized in monolayer 1T-TaSe$_2$ and 1T-TaS$_2$, and there is evidence of such as state from the appearance of incommensurate modulations at wave-vectors given by $2k_F$ expected in a Fermi liquid.~\cite{ruan2021evidence} Further evidence comes from Kondo screening of adsorbed magnetic impurities~\cite{chen2022evidence} We note that recent DMRG calculations find a chiral spin liquid in the vicinity of the Mott transition and not a spinon Fermi surface.~\cite{szasz2020chiral,chen2022quantum} On the other hand, there is no experimental evidence for time reversal breaking in the two systems mentioned above. So we continue to assume that the Fermi surface state is either realized with additional coupling or is a competing state which may arise with carrier doping.

Now we consider doping this spin liquid with carriers which can be electrons or holes. For concreteness we shall use the hole notation. So far the 1T-TaSe$_2$ and 1T-TaS$_2$ cannot be doped. However in the case of the ET salt, it has been possible to introduce Hg chains between the layers, resulting in a doped hole concentration $p$ of 11 \%.~\cite{suzuki2022mott} At the mean-field level, the doped hole will occupy the gapped holon band. On the triangular lattice, the band minimum may occur at the $\Gamma$ point or the $K$ points depending on the sign of $t$. Here we make a further assumption. We consider a strong attraction between the holon and the spinon so they recombine to form physical holes that carry both charge and spin. These holes form a Fermi surface containing $p$ holes if the band bottom is located at $\Gamma$, or they form two Fermi surfaces containing $p/2$ holes each, if the band bottoms are at the zone corner $K$ points. We shall refer to such bands as hole bands. Alternatively, there may be an additional band which happens to be located below the chargon gap, and the doped holes enter that band to form conventional hole pockets. There is in fact evidence for such a band in 1T-TaSe$_2$~\cite{chen2020strong}. The situation is illustrated in Fig.1. The spinon band is half filled and is almost circular. For concreteness the hole band is assumed to be at the $\Gamma$ point and has area corresponding to $p$ spinful holes.

Note that the hole band containing $p$ holes violates Luttinger theorem which states that the Fermi surfaces should enclose a total of $1-p$ fermions. This situation has been dubbed FL$^*$ by Senthil, Sachdev and Vojta.~\cite{senthil2003fractionalized} They explained that one can get around the non-perturbative proof of Luttinger theorem~\cite{oshikawa2000topological} if the spinon sector is in a gapped topological state with degenerate ground states upon flux insertion in a torus geometry. In our case, the spinon sector is gapless and  has massive degeneracy. As a result the non-perturbative proof which relies on returning to the same ground state after a flux insertion in a torus also fails.~\cite{senthil2004weak}. While topology does not play a role, we follow them and denote this as an FL$^*$ state. 
\begin{figure}[h]
\includegraphics[width=0.45\textwidth]{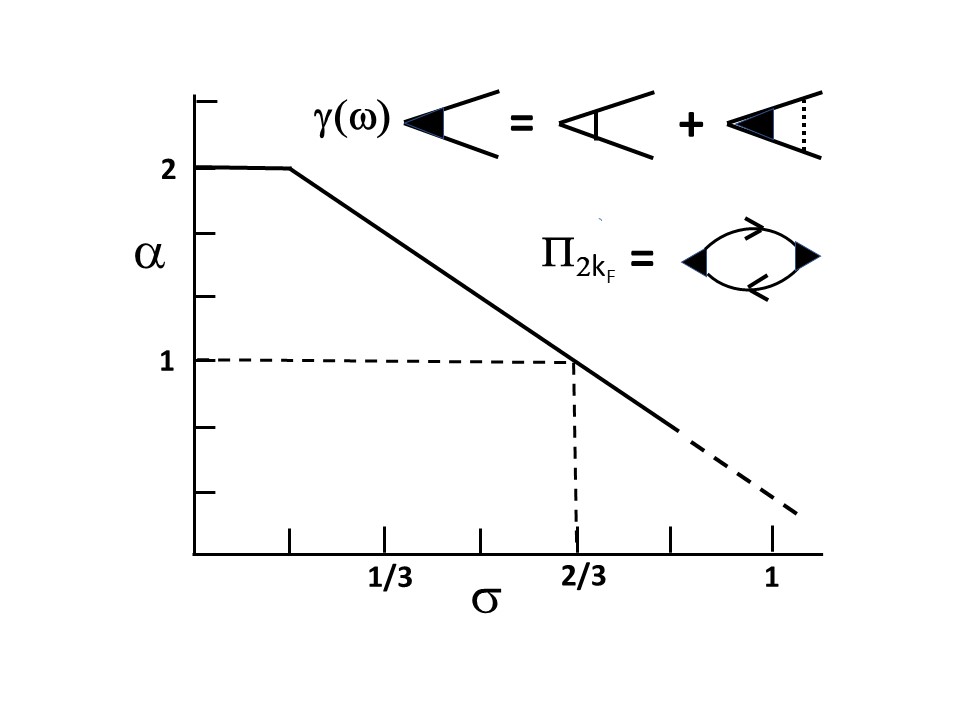}
\caption{A plot of the exponent $\alpha$ which characterizes the resistivity going as $T^{\alpha}$ vs the exponent $\sigma$. Dashed line indicates the breakdown of the linear relation given by Eq.\ref{alpha} for $\sigma >5/6$. The Fermi liquid behave $\alpha=2$ is recovered for $\sigma<1/6$ and linear T resistivity obtains for $\sigma=2/3$, as indicated by the short-dashed line.  $\sigma$ is defined as the divergent exponent of the vertex function $\gamma(\omega) \propto \omega^{-\sigma}$.  As shown in the inset this divergence is due to repeated exchange of the gauge field (dotted line). Also shown is  the diagram for the spinon polarization function $\Pi_{2k_F}$. The intermediate spinon lines denote excitations of spinon particle-hole pairs which are restricted to low energy, hence two factors of the vertex functions appear and contribute the factor $\tilde{\omega}^{-2\sigma}$ in Eq.~\ref{Pi1}. For details see ref.~\cite{altshuler1994low}  } \label{fig2}
\end{figure}
The properties of the hole pocket can be quite conventional, apart from its small Fermi surface area. In contrast, the spinon Fermi surface is strongly coupled to a U(1) gauge field which the holes do not see. This problem has been well studied and one key result is that the self energy goes as $\omega^{2/3}$.~\cite{lee1992gauge} The spinon decay rate exceed $\omega$ which violates Landau's criterion for the existence of quasi-particles. The origin of the strong decay rate is that  the gauge field is strongly overdamped, with a  propagator that goes as $(|\omega|/q+q^2)^{-1}$. The excitation $\omega$ scales as $q^3$ resulting in a copious amount of low energy excitations. For a number of years it was believed that this strong coupled problem can be controlled with a $1/N$ expansion, where $N$ is the number of fermion species. The low energy physics is found to be described by a nontrivial fixed point which is exactly marginal and described by scaling functions.~\cite{altshuler1994low, polchinski1994low} Of particular interest to us is that the polarizibility function for the system comprising spinons coupled to gauge fields, $\Pi(\omega,\vec{Q})$, is singular for $|\vec{Q}|$ near $2k_F$.~\cite{altshuler1994low} Importantly, note that this results in a ring of low energy excitations in momentum space with radius $2k_F$. Unfortunately, upon further scrutiny it was shown by Sung-Sik Lee that the $1/N$ is not controlled even in the large $N$ limit~\cite{lee2009low}. So the nature of the low energy physics is not known even for large $N$, let alone the physical case of $N=2$. Nevertheless, it is still possible that the low energy physics remains to be described by a nontrivial and marginal fixed point. There is support for this possibility by studies involving more expansion parameters.~\cite{mross2010controlled} We will adopt this point of view and assume a scaling form for $\Pi(\omega,\vec{Q})$ which is the same for the large $N$ expansion as given by Altshuler, Ioffe and Millis~\cite{altshuler1994low}, except that we will treat the scaling exponent $\sigma$ as an unknown parameter. ($\sigma$ is the exponent which characterize the divergence of the $2k_F$ vertex function $\gamma$ due to coupling to the gauge fluctuations as shown in Fig 2, namely, $\gamma(\omega) \propto |\omega|^{-\sigma}$.)  
Since $\Pi(\omega,\vec{Q})$ is independent of the direction of $\vec{Q}$,  we introduce the  variable $\Delta \vec{q}=\vec{Q}-2k_F \hat{\vec{Q}}$ to denote the distance of $\vec{Q}$ from the Fermi surface. We write $\Pi(\omega,\vec{Q})$ as $\Pi_{2k_F}(\omega,\Delta \vec{q})$ and denote the imaginary part as
$\Pi^{''}$. We define the scaled variable $\Tilde{q}$ as $\Tilde{q}\hat{\vec{Q}}=\Delta \vec{q}/k_F$, together with
   $\tilde{\omega}=\omega/E_F$. We have,
   \be
\label{Pi1}
\Pi^{''}_{2k_F}(\omega,\Delta \vec{q}) \propto 
\frac{1}{E_F}  \Tilde{\omega}^{2/3-2\sigma} ,\quad \Tilde{\omega}^{2/3} >|\Tilde{q}|
\ee
\be
\label{Pi}
\Pi^{''}_{2k_F}(\omega,\Delta \vec{q}) \propto 
\begin{cases}
\frac{1}{E_F} \Tilde{\omega} |\tilde{q}|^{-1/2-3\sigma}, \quad \Tilde{\omega}^{2/3} <|\Tilde{q}|, \Tilde{q}<0.\\
\frac{1}{E_F} \Tilde{\omega}^{5/3} |\tilde{q}|^{-3/2-3\sigma}, \quad \Tilde{\omega}^{2/3} <|\Tilde{q}|, \Tilde{q}>0.

\end{cases}
\ee
In these equations, the $\sigma $ dependent part comes from the vertex function $\gamma$ and the rest comes from particle-hole  excitations including the self-energy correction. Eq.\ref{Pi1} gives the limit $\Tilde{\omega}^{2/3} >|\Tilde{q}|$ and is given in  ref.~\cite{altshuler1994low}.  The limit $\Tilde{\omega}^{2/3} <|\Tilde{q}|$ is given in Eq.\ref{Pi}. The first line is applicable for $\Tilde{q}<0$ or $|\vec{Q}| < 2k_F$ and apart from the  $
\sigma$ dependent factor, it is the same as the familiar form for free fermions.~\cite{zhang2023free} 
The only difference is that the condition of validity is changed from $\Tilde{\omega} <|\Tilde{q}|$ to $\Tilde{\omega}^{2/3} <|\Tilde{q}|$. The second line in Eq.\ref{Pi} gives the case $|\vec{Q}|> 2k_F$ which is zero for free fermions because we are outside of the particle-hole continuum. In our case there is a finite contribution due to  a self energy which goes as $\omega^{2/3}$.  
Note that Eqs.\ref{Pi1} and \ref{Pi} satisfy the scaling form $\Pi_{2k_F}^{''} \propto \tilde{\omega}^{2/3-2\sigma} 
F(\Tilde{\omega}/\Tilde{q}^{3/2}) $ where $F(x)$ goes to 1 for $x$ small so that $\Tilde{\omega}$ scales as $\Tilde{q}^{3/2}$.  

We note that while Eqs.\ref{Pi1} and \ref{Pi} are often derived assuming a circular Fermi surface, it is generally applicable to any Fermi surface shape as long as opposite $k$ points on the Fermi surface have parallel tangents. In this case $k_F$ is a function of angle $\theta$ and Eqs.\ref{Pi1} and \ref{Pi} remain valid.

\section{Resistivity}
Now we are ready to compute the resistivity of the hole band due to scattering by the soft $2k_F$ mode of the spinon Fermi surface. We note that after shifting by a reciprocal lattice vector $\vec{G}$, the  $2k_F$ vectors are equivalent to a set of vectors $\vec{P}$ centered at the  M point that connect the Fermi surfaces on neigboring extended Brillouin zones, as shown in fig 1a.
If the hole Fermi surface is large enough, these vector can connect points on the hole Fermi surface and give rise to umklapp scattering which relaxes the momentum and current. The condition on the size of the hole Fermi surface is the following. With nearest neighbor hopping the spinon Fermi surface is nearly a circle with radius $k_F \approx 0.375 |\vec{b_1}|$ where $\vec{b_1}$ is the reciprocal lattice vector along $\hat{x}$.~\cite{ruan2021evidence} The length of the shortest vector that connects that two Fermi surfaces is $0.25|\vec{b_1}|$. Let us denote the Fermi momentum of the hole pocket by $p_F$. The condition for umklapp is that $p_F>0.125|\vec{b_1}|$ which is 1/3 of $k_F$ for 1 spinon per unit cell. Hence we conclude that umklapp scattering begins for $p>1/9$ in the case of a single circular hole pocket centered at $\Gamma$. This condition can be relaxed if the Fermi surface deviates from a circle, which is likely the case if the doped holes occupy a separate trivial band. Below we will assume that this condition is satisfied and calculate the hole lifetime due to umklapp scattering. Since the momentum transfer is large and umklapp, the same lifetime will enter the resistivity and all transport phenomena.

Unlike the spinon Fermi surface, the hole Fermi surface is not coupled to the gauge field and does not have the anomalous large self energy and damping rate. The major source of damping at low temperatures come from the umklapp scattering channel under consideration. If this rate is smaller than linear in $T $ or its frequency $\Omega$,  the quasi-particle is well defined and a Fermi liquid description is valid in the Landau sense, even if Luttinger theorem is not obeyed. Here I remark that even if the decay rate ends up with a power law smaller than unity and Landau quasi-particles do not exist, it is still possible to treat the transport problem using Boltzmann equation, as long as the self energy has only frequency dependence and no singular momentum dependence, which is the case here. This was shown by Prange and Kadanoff~\cite{prange1964transport} and their idea have been applied to the fermion coupled to gauge field problem.~\cite{kim1995influence, lee2021low} The idea is that at low frequency, the electron spectral function is a sharp peak in momentum space, and a Fermi surface can be defined by the crossing of this sharp peak across $p_F$. This is familiar in the angle resolved photo-emission spectroscopy (ARPES) literature in that the  momentum distribution curve (MDC) can be sharp while the energy distribution curve (EDC) is broad. This allows us to use the Boltzmann equation approach to calculate the resistivity and the result will remain valid even if the exponent $\alpha$ ends up being less than unity, which will happen in certain parameter range.

In its simplest form, the solution of the Boltzmann equation is just the computation of the scattering rate $1/\tau$ using Fermi's golden rule. As shown in Fig.1b, we label the state on the hole Fermi surface by an angle $\phi$. For simplicity of exposition, we consider the scattering from an initial state at $-\phi_0$ to a final state at the Fermi surface near $\phi_0$ such that they are connected by the shortest umklapp vector which lies along $\hat{x}$. $2k_F$ scattering using longer spanning wave-vector will have similar scattering rate with the same power law. The important point is to note that in general a finite region of $\phi_0$ values will satisfy this umklapp condition and will have similar umklapp scattering rates. Furthermore there are 6 minimal spanning vectors in total which replicate the one shown in fig 1b. Therefore it is quite likely that these regions of initial states will cover the entire Fermi surface. This situation is totally different from the scattering from a critical mode at a finite momentum, such as that due to antiferromagnetic instability. This results in the so called "hot spots" and the problem is that the rapid relaxation is limited to these hot spots and there is a bottleneck to relax momenta from the rest of the Fermi surface.~\cite{hlubina1995resistivity,rosch2000magnetotransport} In order to overcome the bottleneck constraint, one need to introduce disorder scattering which gives a finite resistivity at zero temperature, leading to the difficulties described earlier. 
An important feature of the current model is that the entire Fermi surface or large fragments of it is hot, which allows us to reach the clean limit.

The umklapp scattering rate is given by~\cite{lee2021low}
\be \label{rate}
1/\tau = V_0^2 \int_0 ^\Omega d\omega \int d\phi \ \Pi_{2k_F}^"( \omega, \Delta \vec{q(\phi)})
\ee
where $V_0$ is a short range interaction constant between the  holes and the spinons. Since the imaginary part of $\Pi_{2k_F}$ represents the excitation of a particle-hole pair of spinons, this equation captures the scattering of holes by spinons described in Fig.\ref{fig1}.
The integral is over a final state located at $\phi_0+\phi$ on the Fermi surface and $\vec{Q(\phi)}$ denotes a vector in the direction connecting this point and  the center of the spinon Fermi surface circle which is 
close to $\hat{x}$ for small $\phi$. We are interested in $\vec{Q}(\phi)$ near $2k_F$, so the integration over $\phi$ can be converted to a one dimensional integral over the length of $\Delta \vec{q}$ where $\Delta \vec{q}$ is the deviation of $\vec{Q}$ from the $2k_F$ vector as defined earlier. It is easy to see that $|\Delta \vec{q}| \approx k_F \phi cos(\phi_0)$. Hence the integral over $\phi$ is converted to an integral over $\Tilde{q}/cos(\phi_0)$.
We have 
\be \label{sigma_small}
1/\tau = \frac{V_0^2}{ cos( \phi_0)} \int_0^{\Omega}d\omega \int_{-\Lambda}^\Lambda d\Tilde{q} \ \Pi_{2k_F}^"( \omega, \Delta\vec{ q(\phi)}) 
\ee
where $\Lambda \approx 1$ is an ultra-violet (UV) cut-off in the $\Tilde{q}$ integration. 
We can divide the $\Tilde{q}$ integral into two regions. Region (1) is for $\Tilde{\omega}^{2/3} >|\Tilde{q}|$ as given by  Eq.\ref{Pi1}. The $\Tilde{q}$ integral gives a factor $\Tilde{\omega}^{2/3}$ and we find $1/\tau \propto \Omega^{7/3-2\sigma}$ Region (2) is for $\Tilde{\omega}^{2/3} <|\Tilde{q}|$ as given by  Eq.\ref{Pi}. We consider separately the contributions for $|\vec{Q}|$  less than or greater than $2k_F$ . In the latter case the $\title{q}$ integral is UV convergent since $\sigma$ is  positive, so its value is given by the infra-red cut-off and we find a contribution equal to that of region 1. On the other hand, for $|\vec{Q}|<2k_F$ the integrand is given by the first line in Eq.\ref{Pi} we see that the integral is UV convergent only  for $\sigma>1/6$. 
Hence we find the exponent for the scattering rate $1/\tau \propto \Omega^{\alpha}$ or $T^{\alpha}$, where
\be \label{alpha}
\alpha=7/3-2\sigma , \quad \sigma>1/6
\ee
Note that this result is a consequence of the scaling property for $\Pi_{2k_F}$ described earlier following Eq.\ref{Pi} and does not depend on the values of the exponents  which only serve to control the regime of validity as we next discuss.
From the first line in Eq.\ref{Pi}, for $\sigma<1/6$, the $\Tilde{q}$ integral is infra-red convergent, so its value is given by the UV cutoff $\Lambda$ and is independent of $\tilde{\omega}$. As a result, we find that $\alpha=2$ which dominates over contributions from other regions of the integratioan  and we recover the standard Landau Fermi liquid result. 
On the other hand the spinon self energy is given by $\omega^\alpha$ due to scattering by the holes. When $\alpha<2/3$ the exponent will need to be determined self-consistently. Hence the the validity of Eq.\ref{alpha} is limited to the range $2/3<\alpha<7/3$. This final result is summarized  in Fig 2.

We do not know what the value of $\sigma$ is, but as a matter of principle, this model gives a power law behavior of the resistivity down to the lowest temperature which can include the linear T resistivity. In fact, the linear T case $\alpha=1$ does not play any special role and there is no obvious restriction that $\alpha$ cannot be smaller than unity. 

\section{Discussion}
We have constructed a model based on doping of a spin liquid with a spinon Fermi surface, forming a hole Fermi surface with an area that corresponds to the doped hole concentration $p$. As such it violates Luttinger theorem and belongs to the classification $FL^*$.~\cite{senthil2003fractionalized}
We show that for $p$ large enough, umklapp scattering between the hole Fermi surface and the spinon Fermi surface becomes possible, resulting in a resistivity which goes as $T^\alpha$ where $\alpha$ is given by Eq.\ref{alpha} and can vary over a wide range, depending on the strength of the $2k_F$ singularity of the spinon. In principle $\alpha$ equal to unity or even smaller in possible. In these cases, the Landau criterion for the existence of his quasi-particle is violated and this state can be called a non-Fermi liquid.  Our calculation relies on the formulation of Prange and Kadanoff ~\cite{prange1964transport} who showed that quasi-particles in the Landau sense is not necessary to derive a Boltzmann equation to describe transport. Therefore our result should remain valid for $\alpha$ equal to or less than unity. On the other hand, since the transport properties are based on the Boltzmann equation, other properties such as the Hall constant and magnetoresistance should be conventional. In particular, the magnetoresistance should obey Koehler's rule, which states that the correction to resistivity goes as $(B\tau)^2$. In cuprates a linear in $B$ magnetoresistance is often associated with a linear in $T$ resistivity. It is unlikely that our model or something similar is relevant to the cuprates. The most promising material candidate is the doped organic system where 11\% doping has been achieved and  linear or close to linear in T resistivity which extends over several times of the residual resistivity have been observed over some range of pressure.~\cite{suzuki2022mott} The Hall constant is given by the hole doping in this pressure range, consistent with  small Fermi pockets of total area $p$. Interestingly, 11\% is on the border of applicability of our model, which requires $p>1/9$.

We note that the key ingredient of our model is a critical mode which is soft along a line in momentum space. This gives rise to low energy scattering in finite regions on the Fermi surface which allows us to circumvent the hot spot problem associated with scattering by a mode which is critical at one momentum. We employ a model with a spinon Fermi surface because that represents a critical state which exists over a range in parameter space. Therefore the power law resistivity exists over a range of parameters such as doping or pressure. Furthermore a circular spinon or hole Fermi surface is not needed, because the singularity of $\Pi_{2k_F}$ is a consequence of a two patch model, which rely on the fact that opposite k points in the spinon Fermi surface have parallel tangents.

There has been a lot of discussions concerning  the so-called Planckian bound, $\hbar/\tau<k_BT$.~\cite{zaanen2004temperature} 
At present there appears to be no clear connection between transport lifetime and the Planckian bound.~\cite{hartnoll2022colloquium}
Our model is relatively simple and can serve as a counter-example to the Plankian bound which, unlike phonon scattering, is valid down to zero temperature.

\section*{Acknowledgement} 
I thank Andrey Chubukov and T. Senthil for enlightening discussions and acknowledges support by DOE (USA) office of Basic Sciences Grant No. DE-FG02-03ER46076.

\bibliography{ref}

\newpage

\appendix

\end{document}